\documentclass[fleqn,usenatbib]{mnras}

\usepackage{newtxtext,newtxmath}

\usepackage[T1]{fontenc}
\usepackage{ae,aecompl}

\usepackage{graphicx}	
\usepackage{amsmath}	
\usepackage{amssymb}	

\title[Short title, max. 45 characters]{MNRAS \LaTeXe\ template -- title goes here}

\title[Compton echo from GRBs]{Compton echoes from nearby Gamma-Ray Bursts}

\author[Beniamini et al.]{Paz Beniamini$^{1,2}$\thanks{Email: paz.beniamini@gmail.com}, Dimitrios Giannios$^{3}$, George Younes$^{1,2}$, Alexander J. van der Horst$^{1,2}$, \newauthor Chryssa Kouveliotou$^{1,2}$  \\
	$^{1}$Department of Physics, The George Washington University, Washington, DC 20052, USA \\
	$^2$Astronomy, Physics and Statistics Institute of Sciences (APSIS)\\	
	$^{3}$Department of Physics and Astronomy, Purdue University, 525 Northwestern Avenue, West Lafayette, IN 47907, USA
}

\date{Accepted XXX. Received YYY; in original form ZZZ}

\pubyear{2018}

\begin{document}
	\label{firstpage}
	\pagerange{\pageref{firstpage}--\pageref{lastpage}}
	\maketitle
	
	\begin{abstract}
		The recent discovery of gravitational waves from GW170817, associated with a short Gamma-Ray Burst (GRB) at a distance of $40$Mpc, has demonstrated that short GRBs can occur locally and at a reasonable rate. Furthermore, gravitational waves enable us to detect close by GRBs, even when we are observing at latitudes far from the jet's axis. We consider here Compton echoes, the scattered light from the prompt and afterglow emission. Compton echoes, an as yet undetected counterpart of GRBs, peak in X-rays and maintain a roughly constant flux for hundreds to thousands of years after the burst. Though too faint to be detected in typical cosmological GRBs, a fraction of close by bursts with a sufficiently large energy output in X-rays, and for which the surrounding medium is sufficiently dense, may indeed be observed in this way. The detection of a Compton echo could provide unique insight into the burst properties and the environment's density structure. In particular, it could potentially determine whether or not there was a successful jet that broke through the compact binary merger ejecta. We discuss here the properties and expectations from Compton echoes and suggest methods for detectability.
	\end{abstract}
	
	
	
	
	\section{Introduction}
The recent discovery of gravitational waves, GW170817, associated with a short Gamma-Ray Burst (GRB) at a distance of $40$Mpc \citep{GW170817}, has demonstrated that short GRBs can occur locally and at a reasonable rate. With the projected sensitivity for the next run of joint LIGO / VIRGO observations, neutron star mergers could be detected up to a distance of $\sim 100$Mpc and with a rate of $r_{\rm NSM,obs}=1540_{-1220}^{+3200}\mbox{Gpc}^{-3}\mbox{yr}^{-1}$ corresponding to a detection rate of $0.3-3\mbox{yr}^{-1}$. This detection rate is expected to improve dramatically once LIGO/ VIRGO reach their design sensitivity.

The extremely low luminosity of the $\gamma$-rays from GRB170817 \citep{Goldstein2017}, significantly weaker than typically observed values in GRBs, together with the strong upper limits on the early X-ray afterglow emission have provoked many authors to suggest an off-axis interpretation for the prompt emission, either from a `structured jet' \citep{Kathirgamaraju2018,Fraija2017,Lamb2017} or from a ``cocoon" \citep{Lazzati2017,Gottlieb2017,Kasliwal2017}.

Here we investigate another mechanism that could generate low luminosity X-ray emission from nearby GRBs associated with GW events. 
Photons, predominantly hard X-rays and soft $\gamma$-rays  (up to $\sim 0.5$MeV) from a gamma-ray burst (GRB) exploding in an ionized medium, would produce a reflection signal from Compton scattering on the surrounding electrons. The possibility of such a reflection signal, known as a  ``Compton echo", was first suggested by \cite{Madau2000}. In that work the authors focused on long GRBs, exploding in dense wind environments and calculated the signal due to interaction of the GRB photons with the wind particles. The emission would typically last a few days, before the GRB jet crosses $\sim 10^{16}$cm and the external density starts dropping as $r^{-2}$ leading to a reduction of the luminosity as $t^{-2}$.

Other works (e.g. \cite{Klose1998,Esin2000}) have focused on dust echoes rather than Compton reflection from an ionized ISM. This enables scattering of softer ($\sim 1$keV) X-rays and optical emission and has the potential advantage of a significantly increased scattering cross section. This emission however requires a sufficient source of dust grains to be bright enough and to last any significant period of time.  Another short but enhanced echo signal can accompany long GRBs, if they originate in cold molecular clouds within their host galaxies \citep{Sazonov2003}.

Irrespective of the origin of the emission, it is now becoming clear that the expected number of sources of GWs should increase, and provide triggers for intense EM searches even in cases where no bright prompt EM signal is detected. This motivates us to consider the potential of observing EM signals from neutron star mergers that were previously disregarded due to their inherent dimness because of their large typical distances. In particular this gives a new perspective on the suggestion of Compton echoes, namely the potential of a weaker version of the echoes described by \cite{Madau2000}, associated with reflection of the GRB's prompt and afterglow emission from the host galaxy's interstellar medium, that could be seen up to $\sim 10^4$ yrs after the initial GRB explosion. We focus mainly on the X-ray band, which is likely to hold the most promise for observations. We consider here this possibility, the prospects of its detectability and potential implications for GRB170817.

\section{Luminosity, time-scales and spectrum of the echo}
\label{sec:Lumestimate}
\subsection{Luminosity}
\begin{figure}
	\includegraphics[scale=0.39]{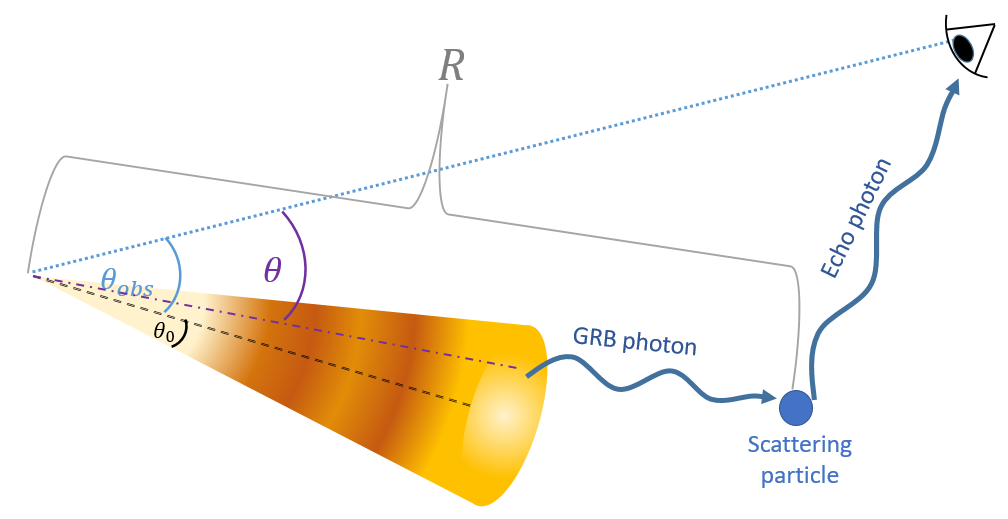}
	\caption{Schematic depiction of Compton echoes of GRBs. The burst is assumed to have an opening angle $\theta_0$ and its core is at a viewing angle of $\theta_{\rm obs}$ from the observer. A GRB photon emitted in the $\theta$ direction is scattered towards the observer by a particle at distance $R$ from the explosion center.}
	\label{fig:illustrate}
\end{figure}
Consider a GRB explosion radiating an energy $E_{\rm GRB}$ in a medium with density $n$. A fraction $\tau$ (where $\tau$ is the Compton optical depth) of the GRB photons are scattered by the surrounding medium. The scattered photons are delayed from the original explosion, due to the longer geometrical path, by a time $t=R(1-cos \theta)/c$, where $\theta$ is the angle between the line of sight and the direction of the reflecting gas as seen from the center of explosion and $R$ is the distance of the scatterer from the center of explosion (see Fig. \ref{fig:illustrate}). Assuming pure Thomson scattering (which is applicable below $\sim 0.5$MeV), the luminosity of the Compton echo is given by (see \cite{Madau2000} for a full derivation)
\begin{equation}
\label{eq:Lecho}
L_{\rm X, echo}=\frac{E_{\rm X, GRB}\tau}{R(1-cos \theta)/c}=\frac{3}{4}\int n \frac{dE_{\rm X, GRB}}{d\Omega}\sigma_T c \frac{1+\cos^2(\theta)}{1-\cos(\theta)}d\Omega
\end{equation}
where $\frac{dE_{\rm X, GRB}}{d\Omega}$ is the GRB energy released within the observed X-ray band and per unit solid angle as a function of $\theta$. Note that we have focused here on 1-10keV X-ray emission, as this band is likely to be the strongest counterpart of the $\gamma$-rays, and the most likely to contribute to the echo. This may be subject to change if for any reason, a significant amount of the intrinsic energy is radiated at a lower energy band.

Since photons reflected at larger distances (emitted over correspondingly longer time-scales) have a proportionally larger probability of being scattered, the overall luminosity remains constant over time, so long as the surrounding density probed by the photons remains constant. The evolution of the Compton echo therefore mimics the density structure of the environment, here assumed to change over galactic ($\sim$kpc) length-scales.

For an on-axis GRB, the X-rays coming directly from the prompt and / or afterglow emission are initially orders of magnitude more powerful than the Compton echo.
However, at later times, the echo can begin to dominate the emission. We denote the viewing angle from the jet axis by $\theta_{\rm obs}$ and the opening angle of the GRB by $\theta_0$.
Assuming that $\theta_0 \ll 1$, the luminosity of the on axis echo, i.e. $\theta_0 \gg \theta_{\rm obs}$ is approximately
\begin{equation}
\label{eq:Lechoon}
L_{\rm X,echo,on} \approx 3 n E_{\rm X,GRB}\sigma_T c \theta_0^{-2}.
\end{equation}

For an off-axis GRB, the prompt emission is beamed away from the observer (although for moderate viewing angles it may still be observable as an X-ray flash, see \citealt{Yamazaki2002}), while the echo can still be seen. Here Eq. \ref{eq:Lecho} simplifies to
\begin{equation}
L_{\rm X,echo} \approx \frac{3}{4} n E_{\rm X,GRB}\sigma_T c \frac{1+\cos^2\theta_{\rm obs}}{1-\cos \theta_{\rm obs}}.
\end{equation}
Assuming $ \theta_{\rm obs}\ll 1$ we obtain
\begin{equation}
\label{eq:Lechooff}
L_{\rm X,echo}\approx 7\times 10^{37}n_{-1}\bigg( \frac{E_{\rm X,GRB}}{10^{50}\mbox{erg}}\bigg) \bigg( \frac{\theta_{\rm obs}}{0.1}\bigg)^{-2} \mbox{erg s}^{-1}
\end{equation}
where $n_{-1}\equiv \frac{n}{0.1\mbox{cm}^{-3}}$.
This is low compared to regular luminosities associated with GRBs, but becomes detectable if the GRB is sufficiently close. The emission would be easiest to detect for slightly off-axis GRBs (where the echo is still strong but the prompt and afterglow emission are weak), or for on-axis GRBs, after the forward shock X-rays have subsided. Indeed this luminosity is comparable to the detected luminosity in the 1-10 keV X-ray counterpart of GW170817 (see discussion in \S \ref{sec:GRB170817}).

\subsection{Time-scale and total energy}
Since short GRBs may often originate from `dead' galaxies, in areas that are not actively star forming \citep{Berger2014}, this emission may in principle last until the GRB photons cross their entire host galaxy. For a galaxy similar to NGC4993, the host of GW170817, with a half light radius of $R_{1/2}=13$kpc, this corresponds to a maximum time-scale of
\begin{equation}
\label{eq:timescale}
t_{\rm max}=\frac{R_{1/2}(1-\cos \theta_{\rm obs})}{c}\approx 2\times 10^3 \bigg(\frac{R_{1/2}}{10\mbox{kpc}}\bigg)\bigg( \frac{\theta_{\rm obs}}{0.1}\bigg)^{2} \mbox{yrs}
\end{equation}
By that time, the total energy carried by the Compton echo is simply:
\begin{equation}
E_{\rm X,echo}=\tau(R_{1/2}) E_{\rm X,GRB}=2\times 10^{47} n_{-1} \bigg( \frac{E_{\rm X,GRB}}{10^{50}\mbox{erg}}\bigg) \bigg(\frac{R_{1/2}}{10\mbox{kpc}}\bigg)\mbox{ erg}.
\end{equation}

\subsection{Potential echo sources}
The strength of the Compton echo depends on $E_{\rm X,GRB}$, the total energy released by the GRB in the the observed X-ray band. This may have different contributions:
\begin{enumerate}
	\item {\it The prompt phase} - For short GRBs the total energy radiated during the prompt phase is typically $10^{49}-10^{50}$erg \citep{Berger2014}. Their spectra peak at $E_p\approx 400$keV, and their average low energy spectral slope is $\alpha=-0.4$ (\citealt{Ghirlanda2009}; $dN/d\nu \propto \nu^{\alpha}$). Assuming that this slope can be extended down to $\lesssim 10$keV where it can be observed by X-ray telescopes, we can expect the prompt GRB to typically contribute to the X-ray echo an energy
	\begin{eqnarray}
	&E_{\rm 1-10 keV, prompt}=E_{\rm prompt} \bigg(\frac{10\mbox{keV}}{E_p}\bigg)^{\alpha+2}  \nonumber \\ & \approx 3\times 10^{47} E_{\rm prompt,50}E_{p,\rm 0.4MeV}^{\frac{\alpha+2}{1.6}}\mbox{ erg}.
	\end{eqnarray}
	where $E_{\rm prompt,50}=E_{\rm prompt}/10^{50}$erg, $E_{p,\rm 0.4MeV}=E_p/0.4$MeV.

	\item {\it The early forward shock afterglow phase} - This phase of the emission depends of course on the existence of a successful jet that breaks out and interacts with the external medium, a topic we return to in \S \ref{sec:GRB170817}. We consider first the `optimal' case for the forward shock afterglow's contribution to the energy released in X-rays. The afterglow emission peaks when the GRB ejecta reaches the deceleration radius. The time at which this happens depends on the blast wave's energy and initial bulk Lorentz factor and on the density of the surrounding medium. It typically occurs a few tens of seconds after the GRB trigger. At that time the electrons accelerated by the forward external shock would still be globally fast cooling ($\nu_m>\nu_c$, where $\nu_m$ is the typical frequency of synchrotron emission from the un-cooled electron distribution and $\nu_c$ is the typical frequency of synchrotron emission from electrons that cool on the dynamical time-scale), and emitting efficiently, only if
	\begin{equation}
	\label{eq:fastcool}
	t_{dec}<9 E_{\rm k,iso,52} n_{-1} \epsilon_{B,-2} \epsilon_{e,-1}^3(1+z) \mbox{sec}
	\end{equation}
	where we have taken into account a possible reduction of the cooling frequency due to IC cooling \citep{SariEsin2001,Zou2009,Beniamini2015}. In the last expression, $\epsilon_{e,-1}\equiv\epsilon_e/0.1$ and $\epsilon_{B,-2}\equiv \epsilon_B/0.01$ correspond to the fraction of shocked energy stored in electrons and magnetic fields respectively, $E_{\rm k,iso,52}\equiv E_{\rm k,iso}/10^{52}$erg is the isotropic equivalent kinetic energy of the blast wave (that could be 5-10 times larger than the energy released in the $\gamma$-rays, see \citealt{Fan2006,Beniamini2015,Beniamini2016}) and $z$ is the cosmological redshift. We have also assumed that $\epsilon_B\leq \epsilon_e$ and for the sake of clarity have taken the slope of the electrons' energy spectrum to be $p=2.5$. As $\nu_m$ decreases more rapidly than $\nu_c$, the former will eventually cross $\nu_c$, after which the electrons become slow cooling, and no longer radiate synchrotron efficiently. The two frequencies meet at (for $\epsilon_B\leq \epsilon_e,p=2.5$):
	\begin{equation}
	\label{eq:cross}
	\nu_{\rm cross}=6\times 10^{17}E_{\rm k,iso,52}^{-1} n_{-1}^{-3/2} \epsilon_{B,-2}^{-1} \epsilon_{e,-1}^{-5/2}(1+z)^{-1} \mbox{Hz}
	\end{equation}
	If the condition given by Eq. \ref{eq:fastcool} is fulfilled and if $\nu_{\rm cross}$ resides within or below the observed X-ray band (which as seen by equation \ref{eq:cross} is highly dependent on the unknown burst parameters), then the energy output in the X-ray band is maximized and is dominated by the flux released at $\nu_m$ as it sweeps through the X-ray band. $E_{\rm 1-10 keV, aft}$ is then proportional to $\nu_m F_{\nu_m}$ and for $\epsilon_B\leq \epsilon_e, p=2.5$ we get:
	\begin{equation}
	E_{\rm 1-10 keV, aft}= 3\times 10^{48}E_{\rm k,50} \epsilon_{e,-1}^{1/2}\epsilon_{B,-2}^{1/2}\mbox{erg}
	\end{equation}
	where $E_{k}\equiv (\theta_0^2/4)E_{\rm k,iso}$ is now the collimated corrected kinetic energy. This expression should be seen as an upper limit on the afterglow's contribution in the X-ray band. If $\nu_m$ crosses the cooling frequency at a frequency above the X-ray band or if the transition to slow cooling occurs earlier than the deceleration time, then the actual energy released in the X-rays would be decreased. In the former case, the X-ray energy output peaks at $t_{\rm c,X}$ which is the time at which $\nu_c$ passes through the X-ray band and is given by (for $\epsilon_B\leq \epsilon_e, p=2.5$)
	\begin{eqnarray}
	& E_{\rm 1-10 keV, aft}\propto \nu_c L_{\nu_c}(t_{\rm c,X})t_{\rm c,X} \\ \nonumber &=2\times 10^{47}E_{\rm k,50}^{-1/2} n_{-1}^{-9/4}\epsilon_{e,-1}^{-9/2}\epsilon_{B,-2}^{1/4}(1+z)^{3/4}\mbox{erg}
	\end{eqnarray}   
	In the later case, most of the X-ray energy is released at $t_{\rm dec}$:
	\begin{eqnarray}
	& E_{\rm 1-10 keV, aft}\propto \nu_c L_{\nu_c}(t_{\rm dec})t_{\rm dec} \\ \nonumber &= 7\times 10^{47} E_{\rm k,50}^{5/6} n_{-1}^{-1/3}\epsilon_{e,-1}^{-3/2}\epsilon_{B,-2}^{3/2}(1+z)^{3/4}\Gamma_2^{-2/3}\bigg(\frac{\theta_0}{0.1}\bigg)^{-8/3}\mbox{erg}
	\end{eqnarray}
	where $\Gamma_2\equiv \Gamma/100$ denotes the initial value of the bulk Lorentz factor.
	\item {\it X-ray flares} - Occasionally, X-ray flares are emitted during the afterglow phase of GRBs \citep{Burrows2005}. The average energy release by X-ray flares is $\sim 2.6$ times smaller than the total GRB X-ray energy release \citep{Margutti2013}. However some early time ($\sim 30$sec) flares, contribute a significant amount towards the total X-ray energy. \cite{Margutti2013} report the isotropic equivalent X-ray energy released in the detected sample of flares. Considering short GRBs only and scaling with the opening angle of the GRB, the X-ray energy release is
	\begin{equation}
	E_{\rm 1-10 keV, flare}= 10^{47}\bigg(\frac{\theta_0}{0.1}\bigg)^2 \mbox{erg}
	\end{equation}
	
	\item {\it Extended emission} - Some GRBs have been shown to have an extended component of soft $\gamma$-rays lasting tens of seconds after the initial hard prompt spike \citep{Lazzati2001,Gehrels2006}. The fraction of short GRBs with this emission is estimated at $\sim 0.15-0.25$ \citep{Norris2010,Berger2014}. When this emission is present, the ratio of energies released in the extended emission (EE) and initial spike vary in the range $E_{\rm EE}/E_{\rm spike}\approx 0.2-34$ \citep{Kaneko2015} with a median value of $\sim 1.7$. The spectral slope of this emission is significantly softer than that of the initial spike, with typical values around $dN/d\nu \propto \nu^{-2}$ \citep{Kaneko2015}, implying an energy in the 1-10keV band which is comparable to the total energy to their observed $\gamma$-rays.
	
\end{enumerate}

From all the above we conclude that for typical short GRBs, the energy released in X-rays is $5\times 10^{46} \mbox{erg} \lesssim E_{\rm X,GRB}\lesssim 5\times 10^{48} \mbox{erg}$. Specifically, in $\sim 1\%$ of short bursts, $E_{\rm X,GRB}$ could be larger than $10^{50}$erg resulting in a significant Compton echo. 

The spectrum of the echo mimics the spectrum of the original X-ray photons. In principle this provides an observational test for the echo scenario (although one that may be practically very challenging, see \S \ref{sec:GRB170817}), and a way of distinguishing between the different possible contributions. Given that Thomson scattering is to a good approximation elastic below $\sim 0.5$MeV, the Compton echo signal can extend to arbitrarily low frequencies. However, the contribution to different energy bands may be dominated by photons produced by the GRB at different times (for instance, the radio flux in GRBs often peaks weeks after the trigger; See \citealt{Chandra2012}). Therefore, provided there is a sufficiently large energy source, Compton echoes may extend down to optical or even radio. In particular, in case the forward external shock dominates the X-ray contribution, a portion of the echo may be observable in the optical and /or radio bands. As the GRB X-ray emission is expected to dominate over the other observational bands, we focus chiefly on this energy band in what follows.

\section{Echo luminosity distribution}
\label{sec:MonteCarlo}
We consider here some typical ranges for the parameters describing the X-ray properties of short GRBs and their surrounding environments, and estimate the detectability of their Compton X-ray echoes. Since short GRBs have been shown to occur relatively close by, and since the detection of gravitational waves associated with these sources could be used as a trigger for a search for Compton echoes, we focus here on this population of bursts. As will be shown in \S \ref{sec:long}, long GRBs are expected to have significantly stronger echoes, and may indeed be detectable if they occur commonly enough in the local universe.

We apply here a Monte Carlo method: we assign a given distribution to each free parameter, draw $10^4$ random values according to those distributions, and using the formulation presented in \S \ref{sec:Lumestimate} calculate the properties of the corresponding Compton echoes. The total energy contributions to the X-ray band by the prompt, the extended emission and the X-ray flares are well constrained by observations. However, the contribution of X-rays by the forward shock afterglow likely peaks at early times, and may often be missed by X-ray telescopes. For this reason we calculate directly the forward shock using the standard formulation (e.g.  \cite{Sari1998,GS2002}) with corrections taking into account IC cooling \citep{Beniamini2015}.

We list here the parameters assumed in our Monte Carlo modeling. We focus on a homogeneous circumburst medium, suitable for the expected environment of short GRBs. We begin with the prompt GRB. We take a log-normal distribution peaking at $\langle E_{\gamma, \rm iso} \rangle=1.6\times 10^{51}$erg with a standard deviation of $\sigma_{\log_{10} E_{\gamma, \rm iso}}=0.2$ \citep{Fong2015}. We then take a normal distribution for the low energy spectral slope, $\alpha$ peaking at $\langle \alpha \rangle=-0.4$ and with $\sigma_{\alpha}=0.5$ and a log normal distribution for $E_p$ peaking at $400$keV and with $\sigma_{E_p}=0.42$ \citep{Ghirlanda2009}. We assume a log-normal distribution for the initial bulk Lorentz factor, centered around $\Gamma=300$ and with $\sigma_{\Gamma}=0.2$ \citep{Ghirlanda2017}. We also take a log-normal distribution for $\theta_0$ with $\langle \theta_0 \rangle=0.1$ and $\sigma_{\log_{10} \theta_0}=0.2$ \citep{Goldstein2016} as well as a random viewing angle, $\theta_{\rm obs}$ (i.e. a flat distribution in $\cos(\theta_{\rm obs})$). These parameters describe the contribution of the prompt spike to the X-rays. For the flares and extended emission we use the distributions of energetics described in \S \ref{sec:Lumestimate}.

For the forward shock contribution we use the following parameters. We adopt a log-normal distribution for $\epsilon_{\gamma}$ with $\langle \epsilon_{\gamma} \rangle=0.15$ and $\sigma_{\log_{10} \epsilon_{\gamma}}=0.2$ \citep{Nava2014,Beniamini2015,Beniamini2016}.
we also assume a log-normal distribution for $\epsilon_e$, with $\langle \epsilon_e \rangle=0.15$ and $\sigma_{\log_{10} \epsilon_e}=0.2$ \citep{Nava2014,Santana2014,BvdH2017}.
$\epsilon_B$ and $n_0$ are less well constrained as compared to the other parameters. We adopt here log-normal distributions with $\langle \epsilon_B \rangle = 10^{-4}$, $\sigma_{\log_{10} \epsilon_B}=1$, $\langle n \rangle =1\mbox{cm}^{-3}$ and $\sigma_{\log_{10} n_0}=1$ \citep[see also][]{Santana2014,Zhang2015,Beniamini2016}). For comparison we also consider distributions centred around $\epsilon_B= 10^{-2}$ and / or $n=0.1\mbox{cm}^{-3}$ with the same scatter in the (logarithm) of those parameters as above. Finally, we consider a uniform range for $p$ between $2.1-2.7$ \citep{GvdH2014}.

\begin{figure}
	\includegraphics[scale=0.39]{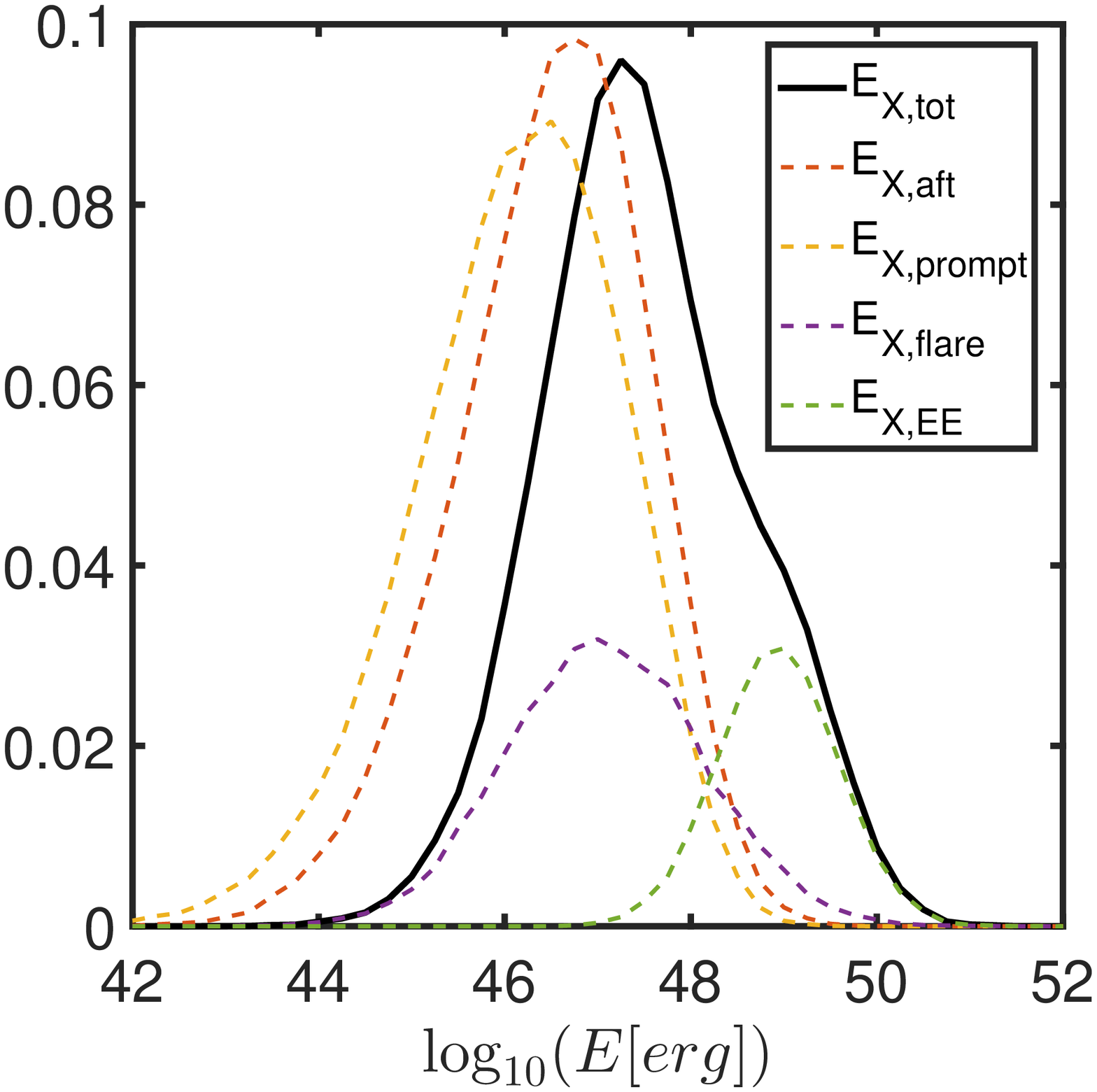}
	\includegraphics[scale=0.39]{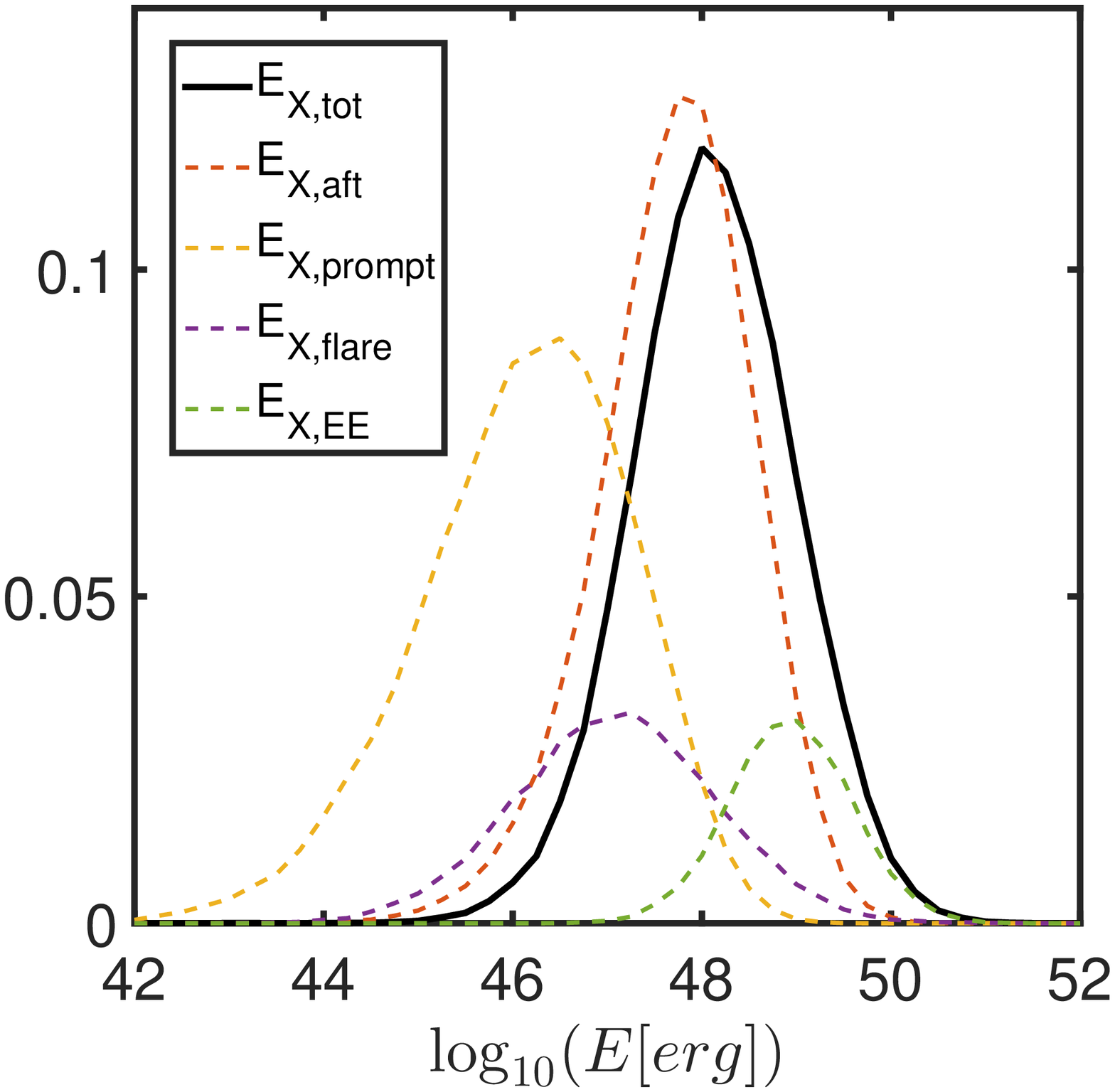}
	\caption{Normalized distribution of energetic contributions (i.e. probability per logarithmic band of energy) to the X-rays by different energy sources in short GRBs: the prompt spike, the forward shock afterglow, X-ray flares and the extended emission. For the forward shock model, the energy is calculated assuming an $\epsilon_B$ distribution that is either centred around $10^{-4}$ (top) or $10^{-2}$ (bottom). These energies depend only weakly on the density of the surrounding medium.}
	\label{fig:energy}
\end{figure}

\begin{figure}
	\includegraphics[scale=0.39]{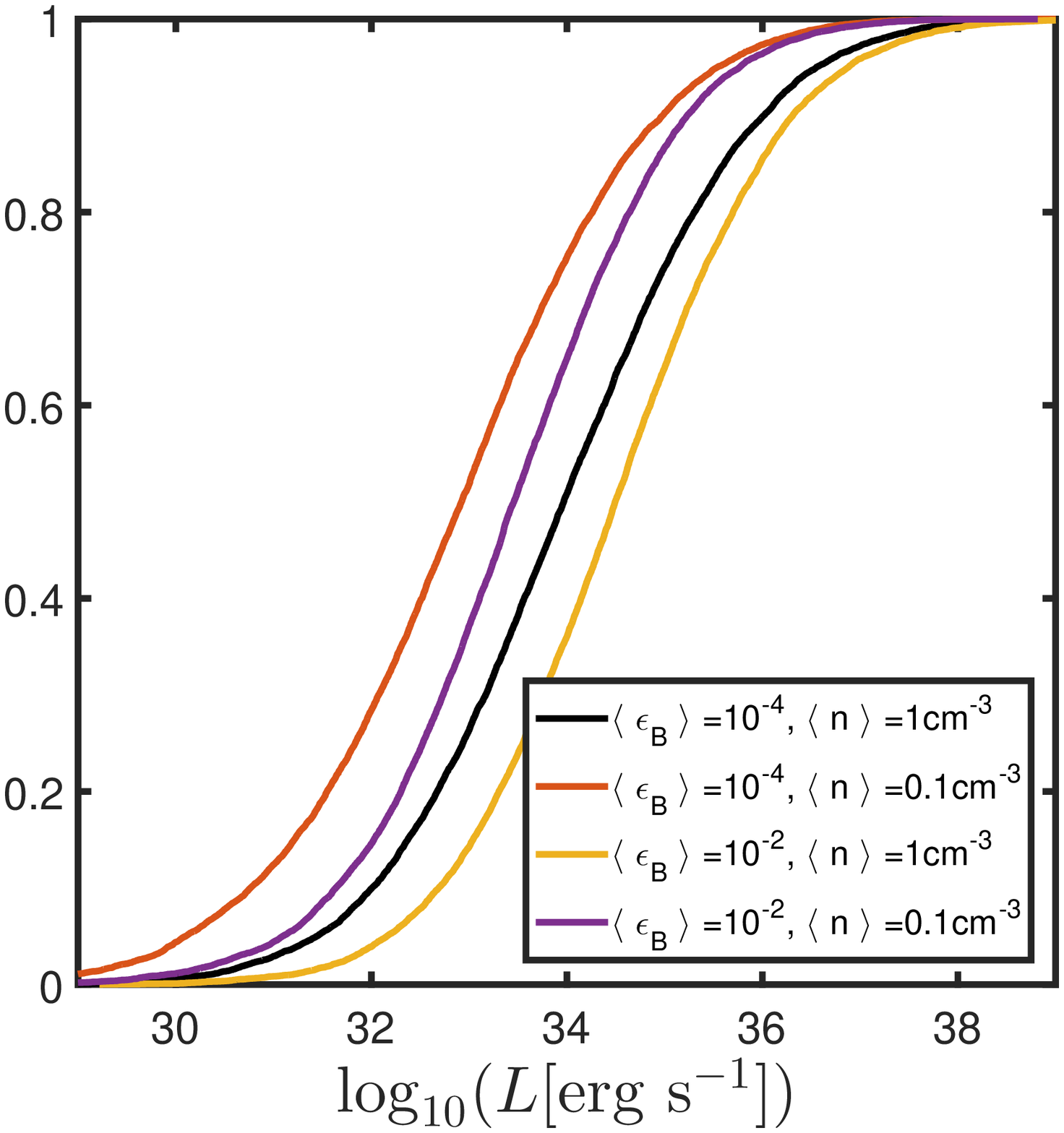}\\
	\includegraphics[scale=0.39]{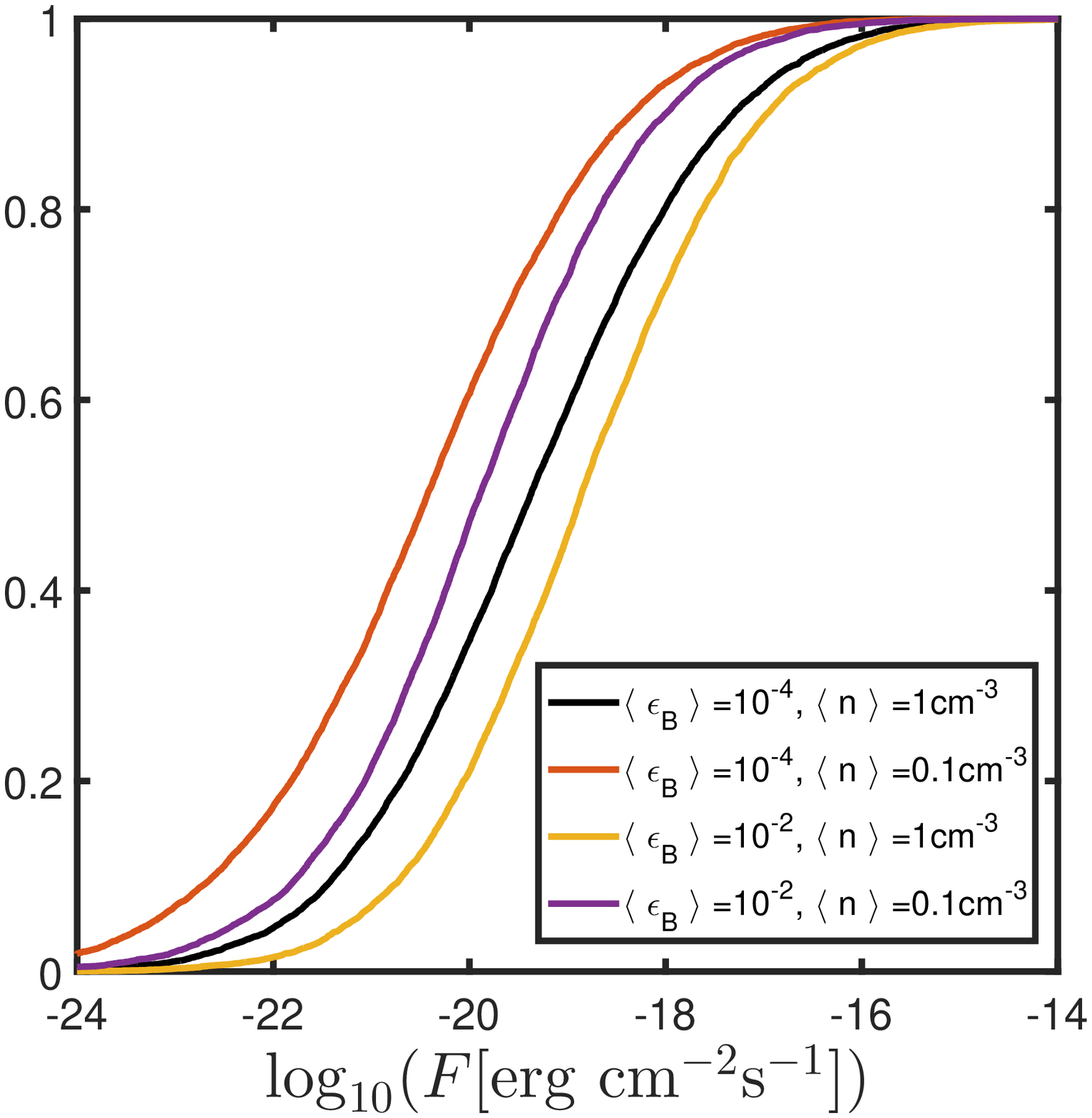}
	\caption{Top: cumulative distribution of total echo luminosities in the 1-10keV band. Bottom: Corresponding distribution of fluxes, assuming events are distributed evenly within a sphere of radius $60$Mpc (corresponding to the approximate limit of detectability of neutron star mergers by gravitational wave detectors).}
	\label{fig:lumflux}
\end{figure}

The distributions of energies of the prompt, forward shock afterglow, X-ray flares and extended emission, resulting from the parameter distributions listed here are shown in Fig. \ref{fig:energy}. Each distribution depicts the probability per logarithmic unit of energy for each of the different components (notice that for the flares and extended emission are assumed to not occur in all bursts, hence the area under those curves is less than unity). For $\langle \epsilon_B \rangle =10^{-4}$, above $\sim 3\times 10^{48}$erg, the energetic contribution is expected to be dominated by the extended emission, while at lower energies the contributions of the external forward shock, X-ray flares and prompt spike become more significant. Increasing the typical values of $\epsilon_B$ to $\langle \epsilon_B \rangle=10^{-2}$, increases the forward shock afterglow contribution by about an order of magnitude, making the forward shock the largest contributor to the total X-ray energy. We also note that changing the typical density from $1\mbox{cm}^{-3}$ to $0.1\mbox{cm}^{-3}$ does not significantly affect the forward shock X-ray energy. The corresponding distribution of echo luminosities can be seen in Fig. \ref{fig:lumflux} for the different magnetization and density models. This figure also depicts the expected fluxes associated with those events, assuming that they are distributed evenly within a sphere of radius $60$Mpc (corresponding to the approximate current limit of detectability of neutron star mergers by gravitational wave detectors). Although the median Compton luminosity is rather low ($\langle L_{\rm X,echo} \rangle=1.1\times 10^{34} \mbox{erg s}^{-1}$), a fraction $\sim 0.01$ of short GRBs should have $L_{\rm X,echo}>3\times 10^{37}\mbox{erg s}^{-1}$ making the prospect of their detectability a promising one.

\section{Implications for nearby long GRBs}
\label{sec:long}
The rate of regular long GRBs in the local universe is not well determined, and likely to be smaller than that of short GRBs \citep{Wanderman2010}. The closest long GRB detected to date, GRB 980425, at a distance of 35.6Mpc, was unfortunately also extremely under-luminous, with a total energy release of `only' $\sim 7\times 10^{47}$erg \citep{Galama1998}.
If, however, regular long GRBs do occur at a reasonable rate in the local universe, their Compton echoes may be detectable in some cases. Furthermore, since their typical energies are considerably larger (and their spectra softer), and given that they may preferentially occur in star forming, and thus denser environments, their echoes could be significantly larger than those of short GRBs.

We adopt here a similar Monte Carlo approach to the one presented in \S \ref{sec:MonteCarlo}. The assumed distribution of parameters are the same as the ones described in that section, with the following exceptions. We take a log-normal distribution peaking at $\langle E_{\gamma, \rm iso} \rangle=1.4\times 10^{53}$erg with a standard deviation of $\sigma_{\log_{10} E_{\gamma, \rm iso}}=0.84$ \citep{Goldstein2016}. We then take a normal distribution for the low energy spectral slope, $\alpha$ peaking at $\langle \alpha \rangle=-0.92$ and with $\sigma_{\alpha}=0.42$ and a log normal distribution for $E_p$ peaking at $200$keV and with $\sigma_{E_p}=0.33$ \citep{Ghirlanda2009}. For the X-ray flares we assume a log-normal distribution of energies peaking at $\langle E_{\rm flare,iso} \rangle=1.8\times 10^{51}$erg with a standard deviation of $\sigma_{\log_{10} E_{\gamma, \rm iso}}=0.97$ \citep{Margutti2013}. Finally, we of course do not consider any extended emission component for long GRBs.
\begin{figure}
	\includegraphics[scale=0.39]{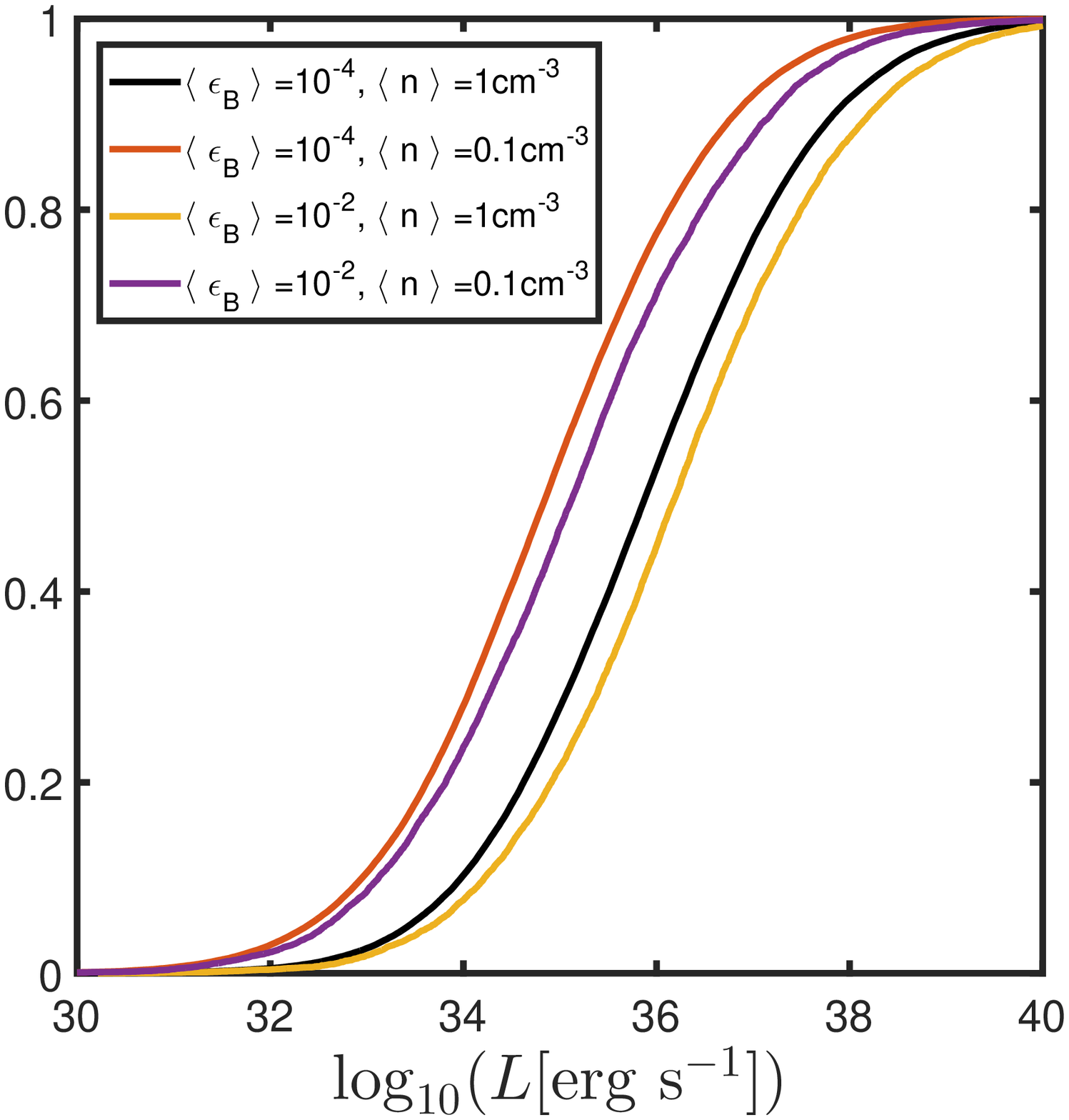}\\
	\includegraphics[scale=0.39]{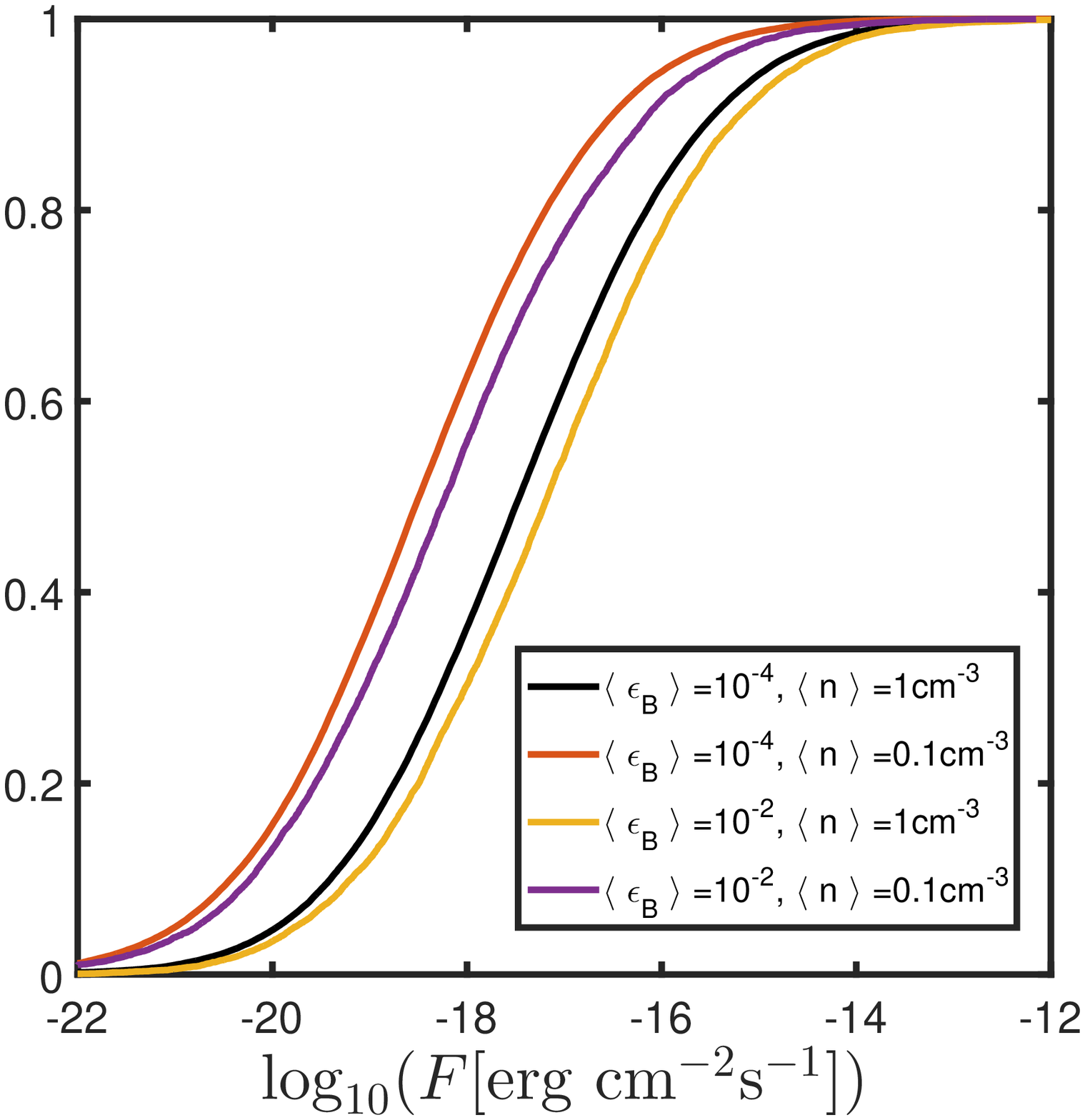}
	\caption{Top: cumulative distribution of total echo luminosities in the 1-10keV band for long GRBs. Bottom: Corresponding distribution of fluxes, assuming events are distributed evenly within a sphere of radius $60$Mpc.}
	\label{fig:lumfluxlong}
\end{figure}

The luminosity and flux distribution of echoes from long GRBs can be seen in Fig. \ref{fig:lumfluxlong}. This time, a fraction $\sim0.15$ of bursts could have an echo luminosity larger than $3\times 10^{37}\mbox{erg s}^{-1}$, and $\sim0.05$ of long GRBs have $L_{\rm X,echo}>3\times 10^{38}\mbox{erg s}^{-1}$.

\section{Observational strategies for detection}
\label{sec:Obs}

\begin{figure*}
	\includegraphics[scale=0.34]{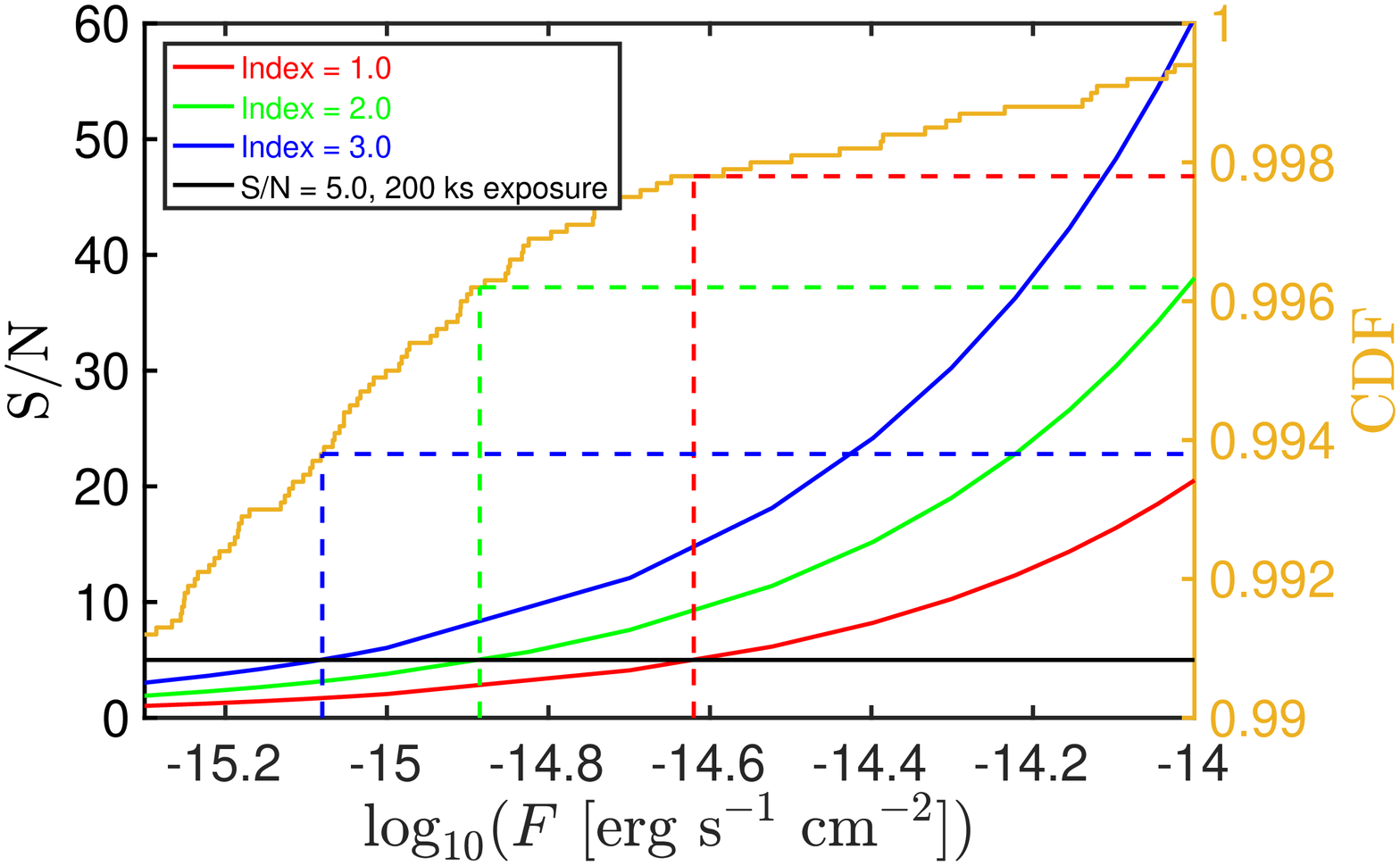}
	\includegraphics[scale=0.34]{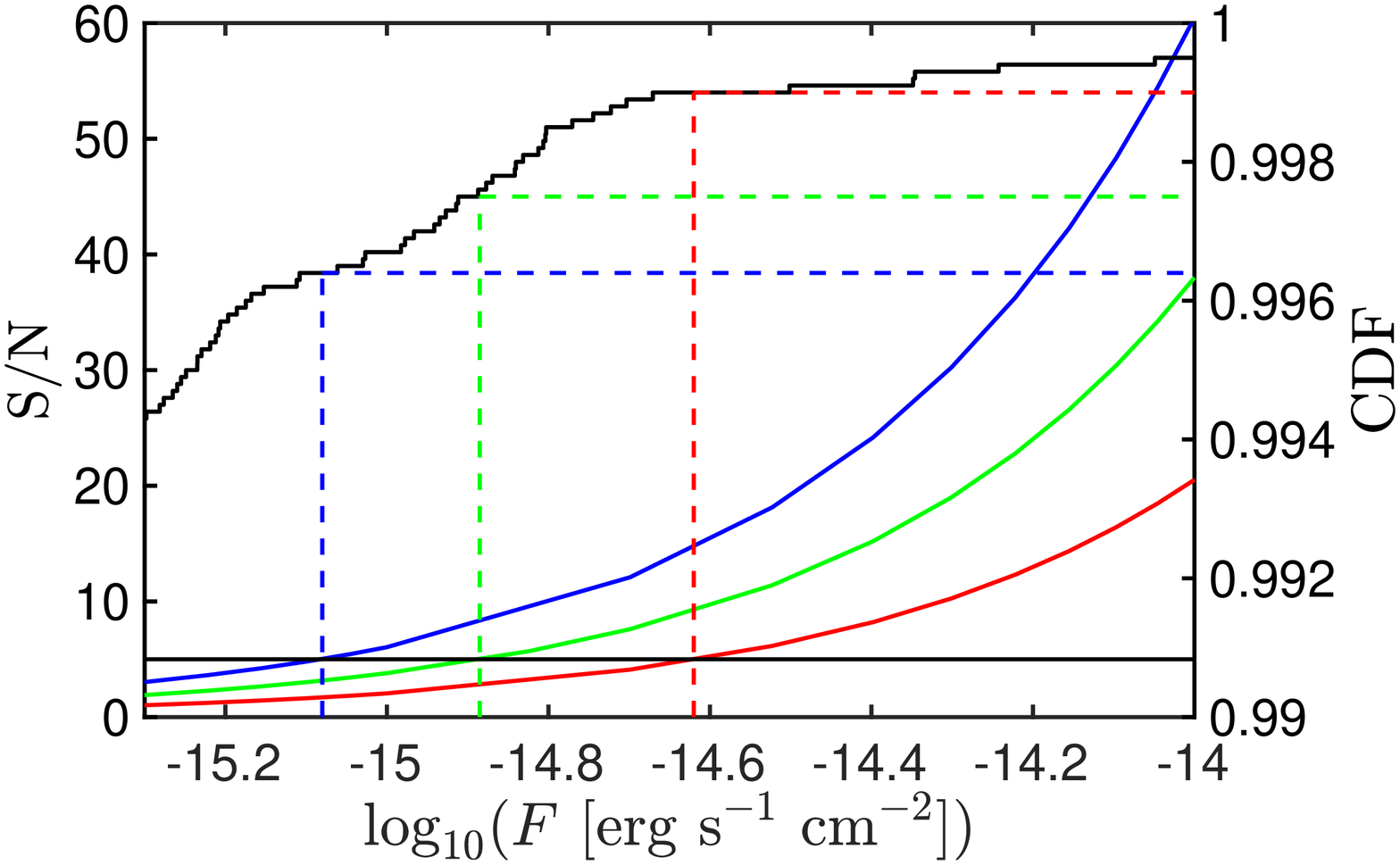}\\
	\includegraphics[scale=0.34]{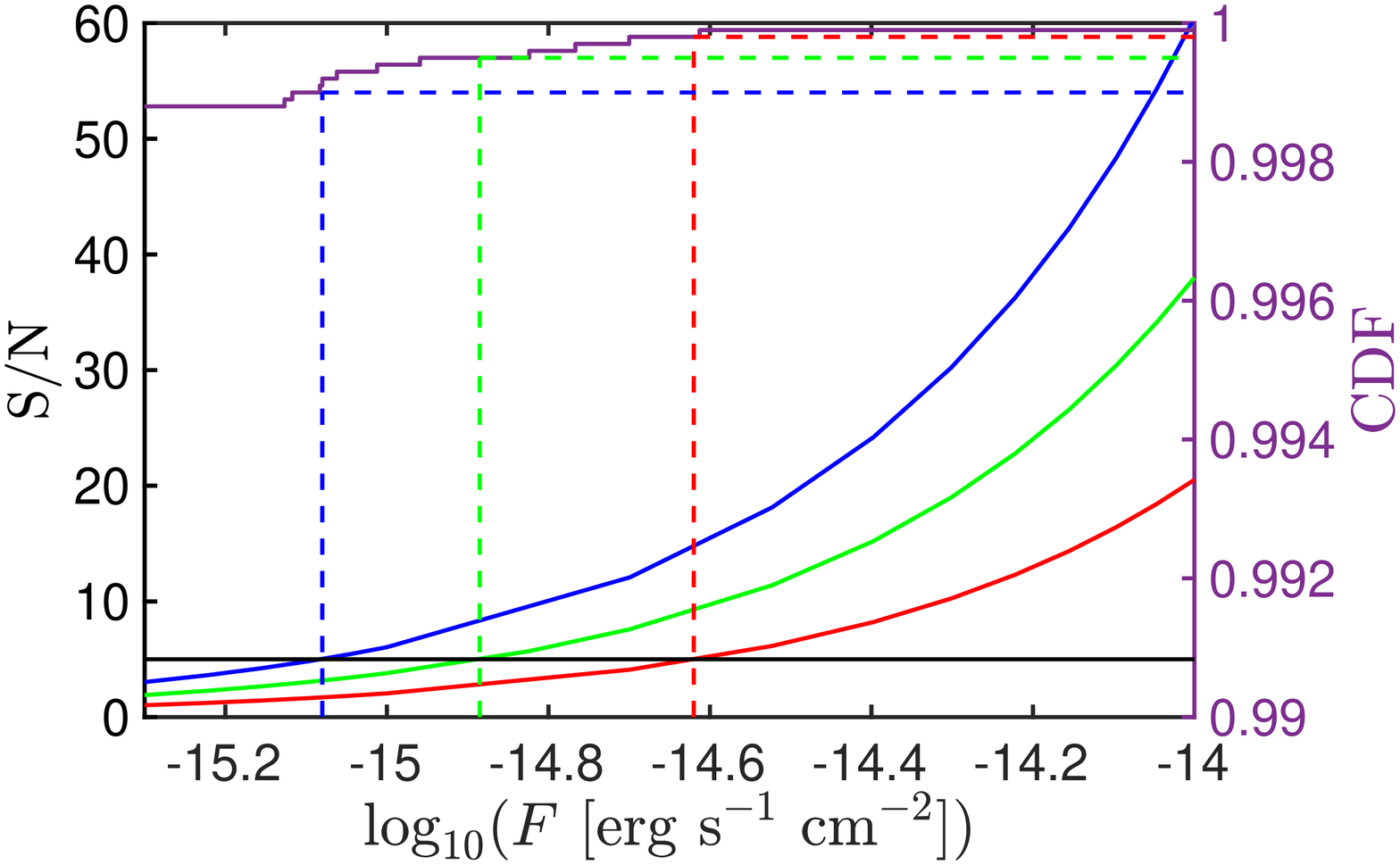}
	\includegraphics[scale=0.34]{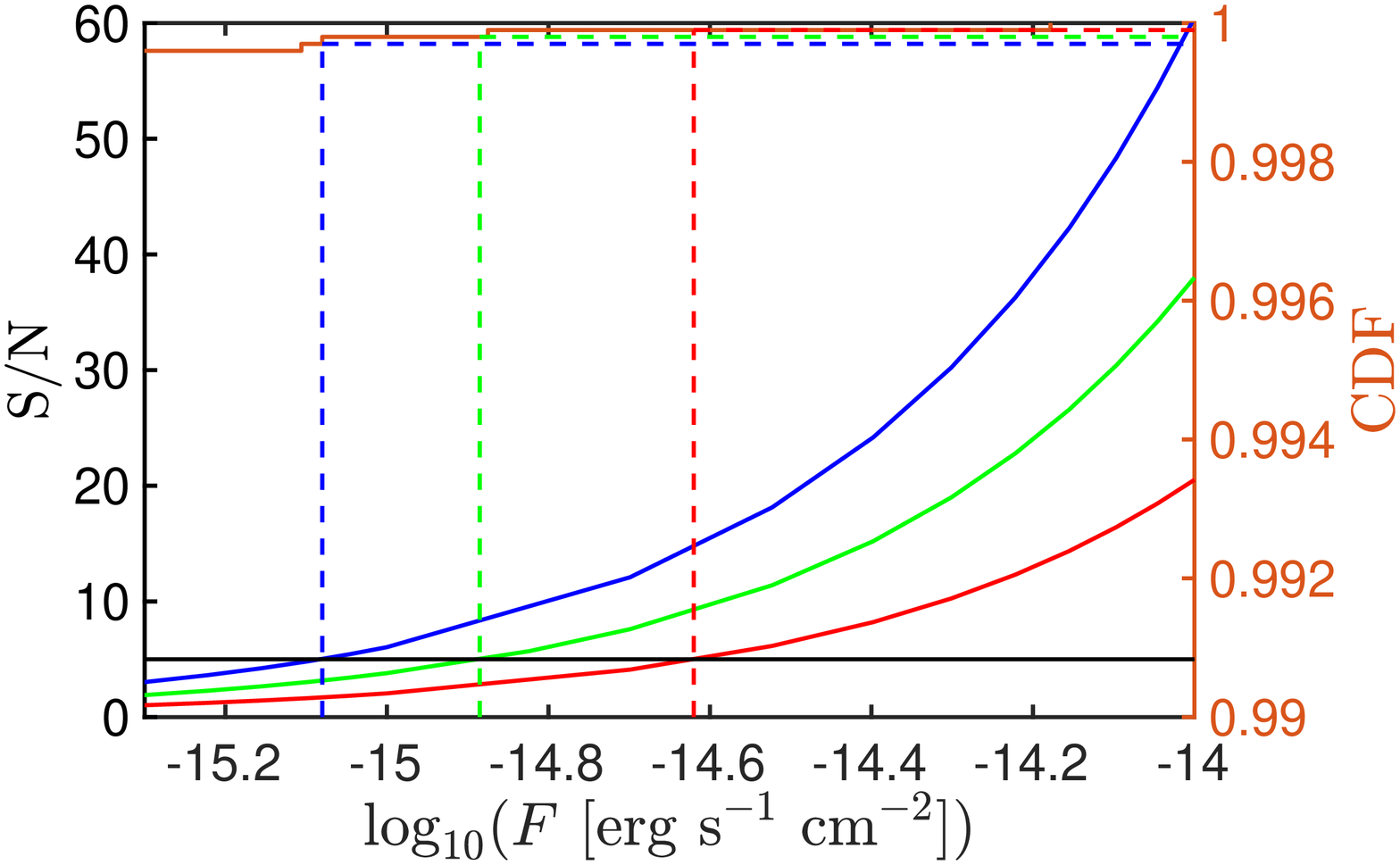}
	\caption{X-ray detectability with {\sl Chandra} of an X-ray echo having a PL X-ray spectrum. In all panels, the blue, green, and red lines show the detectability assuming a PL with a photon index $\Gamma=1.0,~2.0,~{\rm and}~3.0$, respectively. The horizontal black line denotes the minimum detection limit with a 200~ks {\sl Chandra} observation which results in a S/N ratio of 5. The cumulative distribution in each panel refers to the one with the same color shown in the bottom panel of Figure~\ref{fig:lumflux} for different values of $\epsilon_B$ and $n$. See text for more details.}
	\label{fig:echoDetect}
\end{figure*}

Given the low luminosities of Compton echoes, the {\sl Chandra X-ray Observatory} is the most suitable telescope for their detection at the present time. We simulated $10^3$ power-law spectra in the energy range 1-10~keV with fluxes ranging from $5.0\times10^{-16}\mbox{erg cm}^{-2}\mbox{ s}^{-1}$ to $1.0\times10^{-14}\mbox{erg cm}^{-2}\mbox{ s}^{-1}$ for three different photon indices, $\Gamma=1.0,~2.0,~{\rm and}~3.0$. The simulation used {\sl Chandra} response, background, and ancillary files. We assumed the source to be on the aim-point of the ACIS-S3 CCD. Figure~\ref{fig:echoDetect} shows, in all panels, these simulations in red, green, and blue, respectively. The S/N ratio for a detection is given for a 200~ks {\sl Chandra} exposure. The horizontal black line denotes S/N$=5$, which we consider to be a solid detection of the echo. Given the better {\sl Chandra} sensitivity at lower energies, the best sensitivity we could reach with a 200~ks exposure is about $8\times10^{-16}$~erg~s$^{-1}$~cm$^{-2}$ for a photon index of 3.0, while we can reach a flux limit of $2\times10^{-15}$~erg~s$^{-1}$~cm$^{-2}$ for a photon index of 1.0.

Figure~\ref{fig:echoDetect} also shows the cumulative distribution function of the total flux expected from the echo for different parameters $\epsilon_B$ and $n$ (right Y-axis, the colors in each panel correspond to the color of these distribution presented in the bottom panel of Figure~\ref{fig:lumflux}). The largest probabilities of detection are expected for the case with $\epsilon_B=10^{-2}$ and $n=1$~cm$^{-3}$, where somewhere between 2 and 6 in 1000 short GRBs should satisfy our condition of detection. These numbers decrease by about 1 order of magnitude for the case of $\epsilon_B=10^{-4}$ and $n=0.1$~cm$^{-3}$.

Using the observed LIGO rate of neutron star mergers $r_{\rm NSM,obs}=1540_{-1220}^{+3200}\mbox{Gpc}^{-3}\mbox{yr}^{-1}$, and an echo lifetime of $\sim 10^4$ years (see equation \ref{eq:timescale} with isotropically distributed viewing angles, $\theta_{\rm obs}$), we estimate the echo number density at $n_{\rm echo}\sim 0.015_{-0.012}^{+0.032} \mbox{Mpc}^{-3}$. The above {\sl Chandra} detection limits assume that the echo is detected as a point source, i.e., with a 2\arcsec\ PSF, hence, assuming an average apparent expansion of $v_{\rm app}=4.7c$ (the actual value depends on $\theta_{\rm obs}$, see equation \ref{eq:vapp} below) the lifetime for a point source detection is between 2000 and $10^4$ years for a distance ranging from 5 to 60~Mpc, respectively. Beyond the {\sl Chandra} PSF, the small expected fluxes from these echoes become indiscernible from the X-ray background. This translates into an average {\sl Chandra} point source number density of about $0.0015~\mbox{Mpc}^{-3}$, a factor of few smaller than the number density of galaxies that are at least as bright as the Milky Way (and that dominate the baryonic mass in the universe). Considering an average probability of detection with a 200~ks {\sl Chandra} exposure, e.g., 1 in 1000 (Figure~\ref{fig:echoDetect}), the number density for detectable echoes in the nearby universe falls dramatically to $\sim1.5\times10^{-6}$~Mpc$^{-3}$, or $\sim1.5$~Gpc$^{-3}$~yr$^{-1}$ (i.e., 1 in 1000 neutron star mergers should have a detectable echo with a 200~ks {\sl Chandra} observation). The next generation large X-ray telescopes with an order of magnitude better sensitivity than {\sl Chandra}, such as {\sl Lynx} and {\sl Athena}, should improve our chances for the detection of these echoes. The safest strategy for detection is to observe NS mergers one or a few years after the event, when the X-ray afterglow or cocoon emission, has eventually ceased. We note that although the gravitational waves alone are insufficient to uniquely identify a host galaxy, given the brightness of the (roughly isotropic) optical macronova that followed the GW event GW170817 \citep{Smartt2017}, a similar event could be detected up to distances much greater than the current LIGO horizon for BNS mergers (up to 1 Gyr, see \citealt{Metzger2017}) and so is easily within reach of wide-field follow-up telescopes, such as the ZTF \citep{Bellm2014} and the BlackGEM array \citep{Bloemen2016}. EM counterparts in other wavelengths, from radio to X-rays, may also be seen up to at least 60 Mpc (which we consider as the detection horizon in this paper) and may help to localize the event after the GW discovery \citep{Bloom2009,Nakar2011,Kisaka2015} even if we are observing the event off-axis.
A continuous steady X-ray source appearing at the same location of a NS merger after the event, will be strongly supportive of an echo origin, and may be used to glean critical information on the burst and its environment (see \S \ref{sec:GRB170817} for details). Moreover, a high angular resolution instrument combined with high sensitivity might also resolve the early stages of expansion of the echo, which is characterized by its expansion at the speed of light in the local frame, and will be seen as expanding laterally at an apparent transverse velocity of
\begin{equation}
\label{eq:vapp}
v_{\rm app}=\frac{c\sin \theta_{\rm obs}}{1-\cos \theta_{\rm obs}}
\end{equation}
which, depending on $\theta_{\rm obs}$, may appear as sub- or super-luminal transverse motion. Finally, given their large true number density, GRB echoes could be a source of contamination for extragalactic X-ray point source catalogs for the next generation telescopes.

We conclude the detectability discussion with a note regarding radio observations. As mentioned in \S \ref{sec:Lumestimate} the echo emission is expected to peak in X-ray. Even the strongest observed radio peaks (associated with very bright long GRBs), have only released $\lesssim 10^{49}$erg in the radio band \citep{Chandra2012}. This energy is typically released weeks later than the burst trigger, after the time of the jet break \citep{BvdH2017}. The implication is that this emission is no longer significantly beamed, and does not get boosted by the $\max(\theta_0,\theta_{\rm obs})^{-2}$ factor appearing in equations \ref{eq:Lechoon} and \ref{eq:Lechooff} for the X-ray echo. For short GRBs, the total blast-wave energy is considerably lower than for long GRBs (and the densities are likely lower as well) and hence, the energy released in the radio is expected to be at least $\sim 30$ times smaller than for the long bursts. This means that even in the best case scenario, with a release of $\sim 3\times 10^{47}$erg in the radio band, at a distance comparable to GW170817, the flux density at $\sim 10$ GHz, should be roughly $0.3\mu$Jy. Thus, unless an event occurs very close by, the radio echo will be undetectable with current radio telescopes. However, we point out that thanks to very long baseline interferometry, the radio band has an important advantage as compared to X-rays, in terms of its angular resolution, that is better by $\sim 3$ orders of magnitude. This implies that the expansion of the source could be seen within a realistic months to years time-scale, but only with a sensitivity as envisioned for the new generation VLA.

\section{GRB 170817}
\label{sec:GRB170817}
The X-ray counterpart of GRB170817 exhibited an intriguing behaviour. Early X-ray observations provided only upper limits, with $F_X\lesssim 1.7\times 10^{-15}\mbox{erg s}^{-1}\mbox{cm}^{-2}$ between 0.3-10keV \citep{Margutti2017}. The first X-ray detection was made by {\sl Chandra} 9 days after the burst \citep{Troja2017}. The flux was shown to be rising shallowly until the end of the initial observations, at 15 days after the burst, when the flux reached $F_X\approx 5\times 10^{-15}\mbox{erg s}^{-1}\mbox{cm}^{-2}$ \citep{Haggard2017,Margutti2017}. Later observations have shown that the flux continued rising as $t^{0.78\pm 0.05}$ until it reached $F_X\approx 2\times 10^{-14}\mbox{erg s}^{-1}\mbox{cm}^{-2}$ at 109 days \citep{Margutti2018}. The X-ray flux appears to be consistent with being a power-law extrapolation (with $F_{\nu}\propto \nu^{-0.61\pm0.05}$) of the observed radio emission \citep{Hallinan2017,Mooley2017,Margutti2018}. These spectral and temporal tendencies are consistent with both bands originating from a slow cooling synchrotron spectrum (with $\nu_m<\nu_{\rm Radio}<\nu_{X}<\nu_c$) with a continuous source of energy injection. This is consistent with both the structured jet or cocoon interpretation.

An echo origin for the observed X-ray luminosity seems unlikely, unless the true energy of GRB170817 was extremely large and / or the environment very dense. However the former seems unlikely and the latter is in contradiction with the strong limits on the surroundings from the mass of neutral hydrogen $n<0.04\mbox{cm}^{-3}$  \citep{Hallinan2017}. An echo interpretation for the observed X-rays in this burst, seems likely only if there was a sufficient amount of dust for the X-rays to scatter off of (thus greatly increasing the scattering cross section and the associated luminosity). The spectrum of the prompt $\gamma$-rays from GW170817 \citep{Goldstein2017} is very poorly constrained. It is therefore not possible to use the $\gamma$-ray spectrum to confirm or rule out an X-ray echo contribution from the prompt $\gamma$-rays for this burst. Nonetheless, the apparent consistency of the X-rays with the radio data suggests that the X-rays are unrelated with the prompt emission, the X-ray flares or the extended emission. Within the echo interpretation, this leaves only the forward shock afterglow as a possible contributor for the observed X-rays.

Even though the currently observed X-rays are unlikely to originate from an echo, the X-rays originating in either the cocoon or the structured jet interpretation are expected to decay within months (once the emitting material is fully observable and / or the energy injection at the shock has subsided). At that time, an echo may become visible as a persistent source of X-ray emission. In fact, given that an observable echo is likely to be dominated by the forward external shock (see \S \ref{sec:MonteCarlo}), the echo may also be used as a tool for determining the presence or absence of a successful jet that broke out of the merger ejecta. 
\section{Conclusions}
We have considered in this work Compton echoes, a counterpart of GRBs that is expected to peak in the X-ray band and which flux would remain flat for hundreds or thousands of years after the event. Although Compton echoes are weak and cannot be seen for typically observed cosmological GRBs, they may still be detectable for bursts closer than $\sim 60$Mpc. The recent discovery of gravitational waves and the associated short GRB from a neutron star merger, GW170817, that took place at a distance of $\sim 40$Mpc from Earth, demonstrate that such close bursts do indeed occur at a reasonable rate. Furthermore, detection through gravitational wave signals, enables us for the first time to observe GRBs, even if we are looking far from the GRB jet's axis and their emission is not strongly beamed towards us.

The brightness of Compton echoes depends mainly on the total energetic X-ray output of the GRB and on the density of the surrounding medium. Since typically the forward shock afterglow, arising from the interaction of a GRB jet with the external medium, is expected to be the largest contributor to the GRB's X-ray emission, detection of a cocoon may be able to address whether there was a successful jet that broke out of NS-Ns mergers' ejecta, a topic that is currently hotly debated in the literature.
More generally, detection of such echoes would provide us with unique insight into the true (collimated) X-ray energy and opening angle of the GRB jet as well as the density of the surrounding medium, which is also a critical component for modelling of the late-time (days to months) structured jet / cocoon multi-wavelength emission. Furthermore, any fluctuations in the density will be mirrored by the evolution of the echo flux, thus enabling to study in detail the structure of the host galaxy.

Given the low luminosities of these events, {\it Chandra} is the most suitable of the current X-ray missions for detection of Compton echoes.
The safest strategy is to follow-up on NS-NS merger events a year or several years after they are originally observed, when the main X-ray emission, due to the structured jet / cocoon has already subsided. The likelihood of detection is estimated to be small, but not negligible. 
Future X-ray missions will significantly improve the chance of detecting these intriguing counterparts of GRBs.
	\section*{Acknowledgements}
We thank the referee, Pawan Kumar, for his comments on the manuscript.
	

	
	

	\bsp	
	\label{lastpage}
\end{document}